\documentclass[aps,prl,reprint,showpacs,superscriptaddress,twocolumn,10pt]{revtex4-1}%
\usepackage{amsfonts}
\usepackage{mathrsfs}
\usepackage{amsmath}
\usepackage{amssymb}
\usepackage{graphicx}
\usepackage{color}%
\usepackage{mathtools}
\usepackage{booktabs}
\usepackage{mleftright}
\usepackage[colorlinks,linkcolor=blue,citecolor=blue,urlcolor=blue,hyperindex,bookmarks=false,pdfstartview=FitH]{hyperref}

\providecommand{\U}[1]{\protect\rule{.1in}{.1in}}

\newcommand{\figpanel}[2]{\hyperref[#1]{\ref*{#1}(#2)}}

\begin{document}
\title{Giant Atoms in a Synthetic Frequency Dimension}

\author{Lei Du}
\affiliation{Beijing Computational Science Research Center, Beijing 100193, China}
\author{Yan Zhang}
\affiliation{Center for Quantum Sciences and School of Physics, Northeast Normal University, Changchun 130024, China}
\author{Jin-Hui Wu}
\affiliation{Center for Quantum Sciences and School of Physics, Northeast Normal University, Changchun 130024, China}
\author{Anton Frisk Kockum}
\affiliation{Department of Microtechnology and Nanoscience (MC2), Chalmers University of Technology, 412 96 Gothenburg, Sweden}
\author{Yong Li}
\email{liyong@csrc.ac.cn}
\affiliation{Beijing Computational Science Research Center, Beijing 100193, China}
\affiliation{Center for Theoretical Physics and School of Science, Hainan University, Haikou 570228, China}
\affiliation{Synergetic Innovation Center for Quantum Effects and Applications, Hunan Normal University, Changsha 410081, China}

\date{\today }

\begin{abstract}
Giant atoms that interact with real-space waveguides at multiple spatial points have attracted extensive attention due to their unique interference effects. Here we propose a feasible scheme for constructing giant atoms in a synthetic frequency dimension with, e.g., a dynamically modulated superconducting resonator and a tailored three-level artificial atom. Both analytical and numerical calculations show good agreement between our scheme and real-space two-level giant atoms. In particular, the symmetry of the model in momentum space can be broken by tuning the phase of the external field applied on the atom, enabling chiral interactions between the atom and the frequency lattice. We further demonstrate the possibility of simulating cascaded interaction and directional excitation transfer in the frequency dimension by directly extending our model to involve more such effective giant atoms.  
\end{abstract}

\maketitle


\emph{Introduction}.--Giant atoms, which interact with the surrounding environment (waveguides) at multiple points, have attracted rapidly growing interest in the past few years due to various intriguing phenomena arising from them~\cite{fiveyears}. In general, giant atoms can be achieved by coupling (artificial) atoms to propagating fields whose wavelengths are much smaller than the atomic sizes (e.g., surface acoustic waves)~\cite{transmon1,transmon2,transmonA1,transmonA2,transmonA3,transmonA4,transmonA5,transmon3,transmonA6}, or by coupling atoms to meandering waveguides at separated points~\cite{Lamb,braided,EngineerCW}. For such structures, one should naturally consider phase accumulations of photons between different atom-waveguide coupling points, which lead to a series of striking phenomena that are absent for small atoms, such as frequency-dependent Lamb shifts and relaxation rates~\cite{Lamb}, decoherence-free interaction between atoms through the waveguide~\cite{GANori,braided,DFmechanism,AFKchiral}, and photon storage based on bound states~\cite{oscillate,osci2,YuanGA}, to name a few.

On the other hand, the concept of synthetic dimensions has been recently proposed and extensively explored in a variety of physical systems such as photonic structures~\cite{synOL,synOptica,synOptRev,synPRB,LiNbO3PRL,LiNbO3PRB,multiphoton}, cold atoms~\cite{syncold1,syncold2,syncold3,syncold4}, and superconducting circuits~\cite{synsc1,synsc2,synsc3,synsc3,Wilson2021,synSAW}. With synthetic dimensions, it is possible to explore richer physical effects with fewer geometric dimensions. A simple way to create a synthetic dimension is to actively couple modes at different frequencies via dynamical modulations~\cite{synOL,synOptica,synOptRev,synPRB,LiNbO3PRL,LiNbO3PRB,Wilson2021,synSAW}. One can also create synthetic dimensions based on other internal degrees of freedom of photons such as momentum~\cite{SLdwwang,SLOptica} and orbital angular momentum~\cite{OAM1,OAM2,OAM3}. The construction of synthetic dimensions not only enables significant reduction of physical resources, but also provides possibilities for manipulating the relevant degrees of freedom~\cite{synOptRev}.

In this Letter, we demonstrate how to implement giant atoms in a synthetic frequency dimension, where the one-dimensional (1D) frequency lattice acting as a discrete waveguide is achieved with a dynamically modulated superconducting resonator~\cite{synsc3,Wilson2021,synSAW}. We consider a $\Delta$-type artificial atom, where two of the transitions are coupled to different sites of the frequency lattice and the third one is driven by an external field. If the atom supports both a single-photon resonant transition and a two-photon one with large detuning, it can be effectively described as a two-level giant atom with two coupling points in the frequency dimension. We reveal that the effective model can not only simulate some typical effects of real-space giant atoms such as long-lived populations, but can also be used to manipulate light frequency and achieve chiral quantum optical effects in the frequency dimension. 

\begin{figure}[ptb]
\centering
\includegraphics[width=7.5 cm]{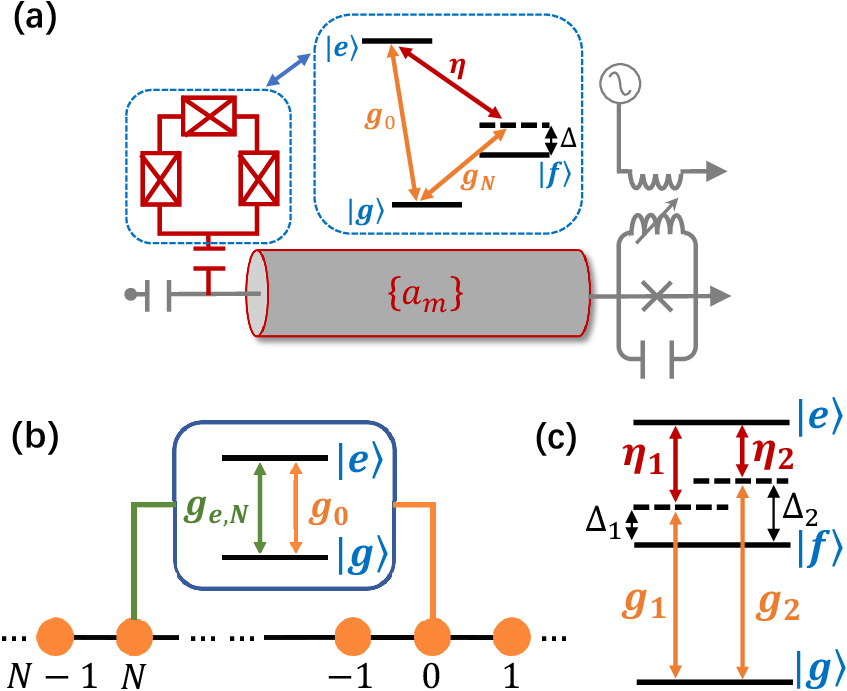}
\caption{(a) Schematic illustration of the model under consideration. The SQUID coupled to the superconducting resonator is modulated in time. A $\Delta$-type artificial atom interacts with two resonant modes of the resonator and is driven by an external field. (b) Effective two-level giant atom after adiabatic elimination. (c) Ladder-type implementation scheme of the effective giant atom.}\label{fig1}
\end{figure}

\emph{Model}.--As shown in Fig.~\figpanel{fig1}{a}, we consider a multimode superconducting transmission line resonator that is connected at one end to a superconducting quantum interference device (SQUID)~\cite{synsc3,Wilson2021,synSAW,ParaAm1,synDelsing,ParaAm2,synShumeiko}. The frequencies of the resonant modes are nearly equally spaced within a specific band~\cite{synsc3}. By modulating the SQUID at a frequency near the free spectral range of this band, coherent couplings between adjacent resonant modes can be introduced and thus a tight-binding frequency lattice is created. The Hamiltonian of the frequency lattice in the interaction picture can be written as (see Sec.~I of the Supplemental Material~\cite{SM} for more details; $\hbar=1$ throughout this Letter)~\cite{synsc3,Wilson2021,synSAW}    
\begin{equation}
H_{r}=\sum_{m}J(a_{m}^{\dag}a_{m+1}+\textrm{H.c.}),
\label{eq1}
\end{equation}
where $a_{m}$ ($a_{m}^{\dag}$) is the annihilation (creation) operator of the $m$th resonant mode with frequency $\omega_{m}$ ($m=0,\pm1,\pm2, ...$); $J$ is the nearest-neighbor coupling strength that is related to the Josephson energy of the SQUID; $\text{H.c.}$ denotes the Hermitian conjugate. In Eq.~(\ref{eq1}), we have assumed that the modulation frequency is resonant with the free spectral range.

Moreover, we consider a $\Delta$-type artificial atom with ground state $|g\rangle$, middle state $|f\rangle$, and excited state $|e\rangle$. Such a structure can be readily achieved with superconducting qubit circuits operated away from the optimal points~\cite{LiuYX2005,coexist,GuReview}, but is not allowed with electric-dipole transitions of natural atoms due to selection rules~\cite{Tannoudji}. We select $|g\rangle$ as the reference so that the energies of $|f\rangle$ and $|e\rangle$ are denoted by $\omega_{f}$ and $\omega_{e}$, respectively. The transitions $|g\rangle\leftrightarrow|e\rangle$ and $|g\rangle\leftrightarrow|f\rangle$ are coupled to resonant modes $a_{0}$ and $a_{N}$ ($N<0$ if $\omega_{f}<\omega_{e}$), respectively, while the transition $|f\rangle\leftrightarrow|e\rangle$ is driven by a coherent external field of amplitude (frequency) $\eta$ ($\omega_{d}$). In the interaction picture and with the rotating-wave approximation, the total Hamiltonian of the system can be written as (see Sec.~II of Ref.~\cite{SM})
\begin{equation}
\begin{split}
H & = \Big(g_{0}a_{0}^{\dag}|g\rangle\langle e|e^{i\Delta_{1}t}+g_{N}a_{N}^{\dag}|g\rangle\langle f|e^{i\Delta_{2}t}\\
& \quad+\eta e^{i\theta}|e\rangle\langle f|e^{i\Delta_{d}t}+\text{H.c.}\Big)+H_{r},
\end{split}
\label{eq2}
\end{equation}
where $g_{0}$ ($g_{N}$) is the coupling strength between mode $a_{0}$ ($a_{N}$) and the transition $|g\rangle\leftrightarrow|e\rangle$ ($|g\rangle\leftrightarrow|f\rangle$), and is assumed to be real for simplicity; $\theta$ is the phase of the external field and determines the global phase of the closed-loop level structure~\cite{LiuYX2005}; $\Delta_{1}=\omega_{0}-\omega_{e}$ ($\Delta_{2}=\omega_{N}-\omega_{f}$) is the detuning between mode $a_{0}$ ($a_{N}$) and the transition $|g\rangle\leftrightarrow|e\rangle$ ($|g\rangle\leftrightarrow|f\rangle$); $\Delta_{d}=\omega_{e}-\omega_{f}-\omega_{d}$ is the detuning between the transition $|f\rangle\leftrightarrow|e\rangle$ and the external field. In the case of $\Delta_{1}=0$ and $\Delta_{2}=\Delta_{d}=\Delta\neq0$, where a single-photon and a two-photon resonant transition coexist, Eq.~(\ref{eq2}) is equivalent to the time-independent Hamiltonian
\begin{equation}
\begin{split}
H'&=H_{r}-\Delta|f\rangle\langle f|+\Big(g_{0}a_{0}^{\dag}|g\rangle\langle e|\\
&\quad+g_{N}a_{N}^{\dag}|g\rangle\langle f|+\eta e^{i\theta}|e\rangle\langle f|+\text{H.c.}\Big).
\end{split}
\label{eq3}
\end{equation}
The state of the system in the single-excitation subspace can be written as
\begin{equation}
|\psi(t)\rangle=\Big[\sum_{m}u_{m}(t)a_{m}^{\dag}+\sum_{\beta=f,e}u_{\beta}(t)|\beta\rangle\langle g|\Big]|0,g\rangle,
\label{eq4}
\end{equation}
with which one can solve the Schr\"{o}dinger equation and obtain
\begin{equation}
\begin{split}
i\dot{u}_{e}&=-i\gamma u_{e}+g_{0}u_{0}+\eta e^{i\theta}u_{f},\\
i\dot{u}_{f}&=-i\gamma u_{f}-\Delta u_{f}+\eta e^{-i\theta}u_{e}+g_{N}u_{N},\\
i\dot{u}_{m}&=J(u_{m+1}+u_{m-1})+g_{0}u_{e}\delta_{m,0}\\
&\quad+g_{N}u_{f}\delta_{m,N},
\end{split}
\label{eq5}
\end{equation}
where $u_{m}$ ($u_{\beta}$) is the probability amplitude of creating a photon in the $m$th resonant mode (of the atom in the state $|\beta\rangle$); $\gamma$ describes the intrinsic dissipation of the atom (assumed to be identical for $|e\rangle$ and $|f\rangle$), which can be much smaller than the other interaction strengths for superconducting qubits~\cite{braided}. Note that the excitation number operator $N_{e}=\sum_{m}a_{m}^{\dag}a_{m}+\sum_{\beta}|\beta\rangle\langle\beta|$ commutes with the Hamiltonian in Eq.~(\ref{eq3}), implying that the total excitation number of our model is conserved.

\emph{Giant-atom effects}.--When $\Delta\gg \{g_{0},g_{N},\eta\}$, $|f\rangle$ can be adiabatically eliminated if it is initially unpopulated~\cite{Stark1993,AdiaElim,YanDong}. If we further assume $\gamma\ll \{g_{0},g_{N},\eta\}$, Eq.~(\ref{eq5}) can be approximately simplified to
\begin{equation}
\begin{split}
i\dot{u}_{e}&=(\Delta_{e,e}-i\gamma)u_{e}+g_{0}u_{0}+g_{e,N}e^{i\theta}u_{N},\\
i\dot{u}_{m}&=J(u_{m+1}+u_{m-1})+(g_{0}u_{e}\delta_{m,0}\\
&\quad+(g_{e,N}e^{-i\theta}u_{e}+\Delta_{e,N}u_{N})\delta_{m,N},
\end{split}
\label{eq6}
\end{equation}
where $\Delta_{e,e}\approx\eta^{2}/\Delta$ and $\Delta_{e,N}\approx g_{N}^{2}/\Delta$ are the effective frequency shifts of $|e\rangle$ and $a_{N}$, respectively; $g_{e,N}\approx g_{N}\eta/\Delta$ is the effective coupling strength between mode $a_{N}$ and the transition $|g\rangle\leftrightarrow|e\rangle$ due to the two-photon process. If $\theta=0$ and $\{\Delta_{e,e},\Delta_{e,N}\}\rightarrow0$, Eq.~(\ref{eq6}) is similar to the dynamic equations of a two-level giant atom coupled to a real-space $1$D lattice at two separated sites~\cite{LonghiGA}, which reads
\begin{equation}
\begin{split}
i\dot{w}_{e}&=-i\gamma w_{e}+\lambda_{0}w_{0}+\lambda_{N}w_{N},\\
i\dot{w}_{m}&=J(w_{m+1}+w_{m-1})+w_{e}(\lambda_{0}\delta_{m,0}+\lambda_{N}\delta_{m,N}).
\end{split}
\label{eq7}
\end{equation}
Here $w_{m}$ and $w_{e}$ are the probability amplitudes of exciting the $m$th site of the lattice and the atom, respectively$; \lambda_{0}$ ($\lambda_{N}$) is the atom-lattice coupling strength at the $0$th ($N$th) site. In view of this, our model is equivalent to a two-level giant atom in the frequency dimension, as shown in Fig.~\figpanel{fig1}{b}. For superconducting qubits, $g_{0}\ll g_{N}$ is allowed such that the coupling strengths $g_{0}$ and $g_{e,N}$ of the effective giant atom can be identical~\cite{GuReview}. It is challenging, however, to simulate such an effective giant atom with a cyclic three-level natural atom (see Sec.~I of Ref.~\cite{SM} for more details).

Before proceeding, we point out that the effective giant atom in the frequency dimension can also be implemented with a ladder-type three-level atom, which is coupled to different sites of the frequency lattice via multiple two-photon resonant transitions, as shown in Fig.~\figpanel{fig1}{c}. The feasibility and advantages of such a scheme are discussed in detail in Sec.~III of Ref.~\cite{SM}. Considering the schemes of creating synthetic dimensions based on optical resonators~\cite{synOL,synOptica,synPRB,synOptRev}, this also shows the possibility of constructing effective giant atoms in the optical regime.

\begin{figure}[ptb]
\centering
\includegraphics[width=8 cm]{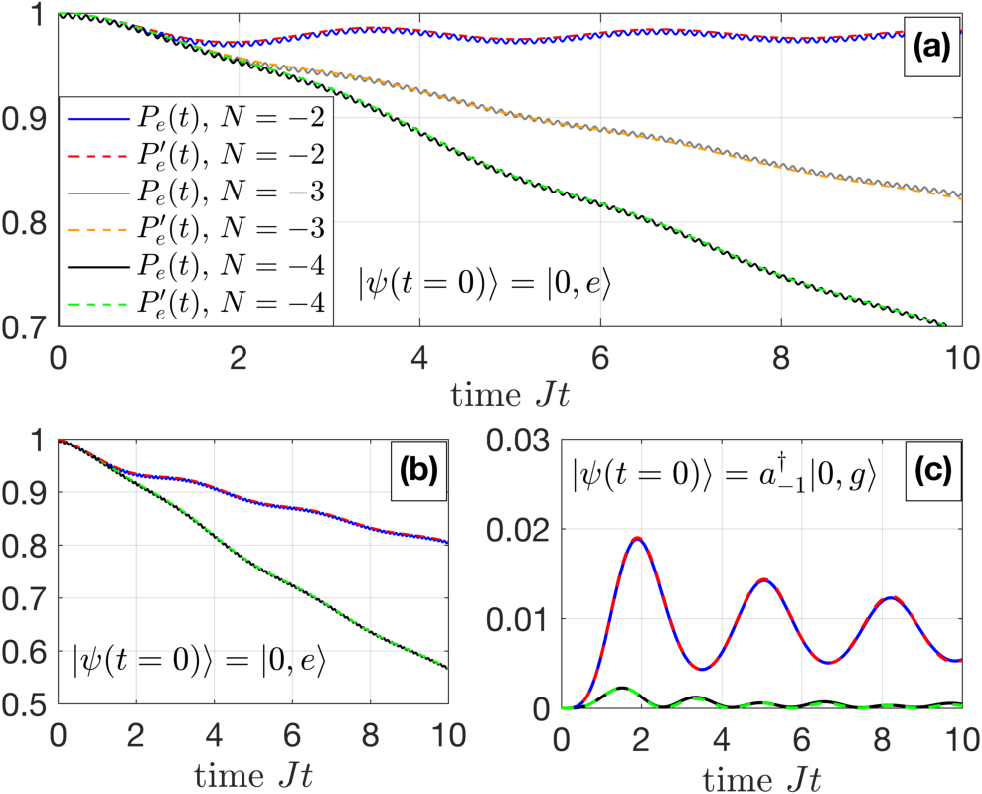}
\caption{Dynamic evolutions of $P_{e}(t)$ and $P_{e}'(t)$ with different values of $N$ and initial states. We assume $\gamma=0$ in (a) and $\gamma=0.01J$ in (b) and (c). Other parameters are $g_{0}=0.1J$, $g_{N}=3J$, $\eta=2J$, $\theta=0$, $\lambda_{0}=\lambda_{N}=0.1J$, and $\Delta=60J$.}\label{fig2}
\end{figure}

We first verify the validity of the effective giant atom in the case of $\theta=0$ by numerically solving Eq.~(\ref{eq5}) with appropriate parameters and comparing the results with those of the real-space correspondence [obtained with Eq.~(\ref{eq7})]. In practice, the frequency dimension typically extends over a limited number of resonant modes where the free spectral range is nearly constant. Therefore we consider $30$ modes (lattice sites) in total~\cite{synSAW} and assume that $|N|$ is much smaller than the lattice length to avoid the boundary effect. Figure~\figpanel{fig2}{a} shows the evolutions of the atomic populations $P_{e}(t)=|u_{e}(t)|^{2}$ and $P_{e}'(t)=|w_{e}(t)|^{2}$  with the initial state $|\psi(t=0)\rangle=|0,e\rangle$ and different values of $N$ in the ideal case of $\gamma=0$. Clearly, the results of the two models can be well fitted if $\lambda_{0}=g_{0}$ and $\lambda_{N}=g_{e,N}$. In particular, long-lived population can be observed if $N=-2$ and $g_{0}=g_{e,N}$, similar to a decoherence-free giant atom in real space~\cite{Lamb,LonghiGA}. Physically, this arises from the destructive interference between the two atom-waveguide coupling paths. Although the two paths correspond to different frequencies, hopping along the frequency lattice causes frequency conversion of the wave such that the giant-atom interference effect can still take place. Moreover, it shows that the decay rate of $P_{e}(t)$ depends on $N$, which reproduces the phase-dependent spontaneous emission of a real-space giant atom. The population evolutions for $\gamma\neq0$ are demonstrated in Fig.~\figpanel{fig2}{b}, which exhibits somewhat accelerated atomic decay. Hereafter, we will always take the intrinsic dissipation of the atom into account.

We also plot in Fig.~\figpanel{fig2}{b} the evolutions of $P_{e}(t)$ and $P_{e}'(t)$ with another initial state $|\psi(t=0)\rangle=a_{-1}^{\dag}|0,g\rangle$, which again show good agreement between these two models. While the atom can hardly be excited if $N=-4$, an oscillating atomic population arises in the case of $N=-2$. This can be understood again from the two-path interference effect. For $N=-4$, the phase accumulations of the photon traveling from its initial position (i.e., site $a_{-1}$) to the atom via the two coupling points are $\phi_{1}=-\pi/2$ and $\phi_{2}=-3\pi/2$~\cite{LonghiGA}, respectively, such that the two excitation paths of the atom interfere destructively. For $N=-2$, however, $\phi_{1}=\phi_{2}=-\pi/2$ leads to constructive interference between the two paths. Because of the weak atom-waveguide couplings, only a portion of the excitation can oscillate between mode $a_{-1}$ and the excited state $|e\rangle$ and the oscillation tends to vanish eventually. We also provide in Sec.~IV of the Supplemental Material~\cite{SM} an effective model of ``giant-small atoms''~\cite{synEIT}, where the initially occupied lattice site is effectively replaced by a vacuum one coupled with an excited small atom, to give a more intuitive and quantitative explanation for the result above.     

\begin{figure}[ptb]
\centering
\includegraphics[width=8 cm]{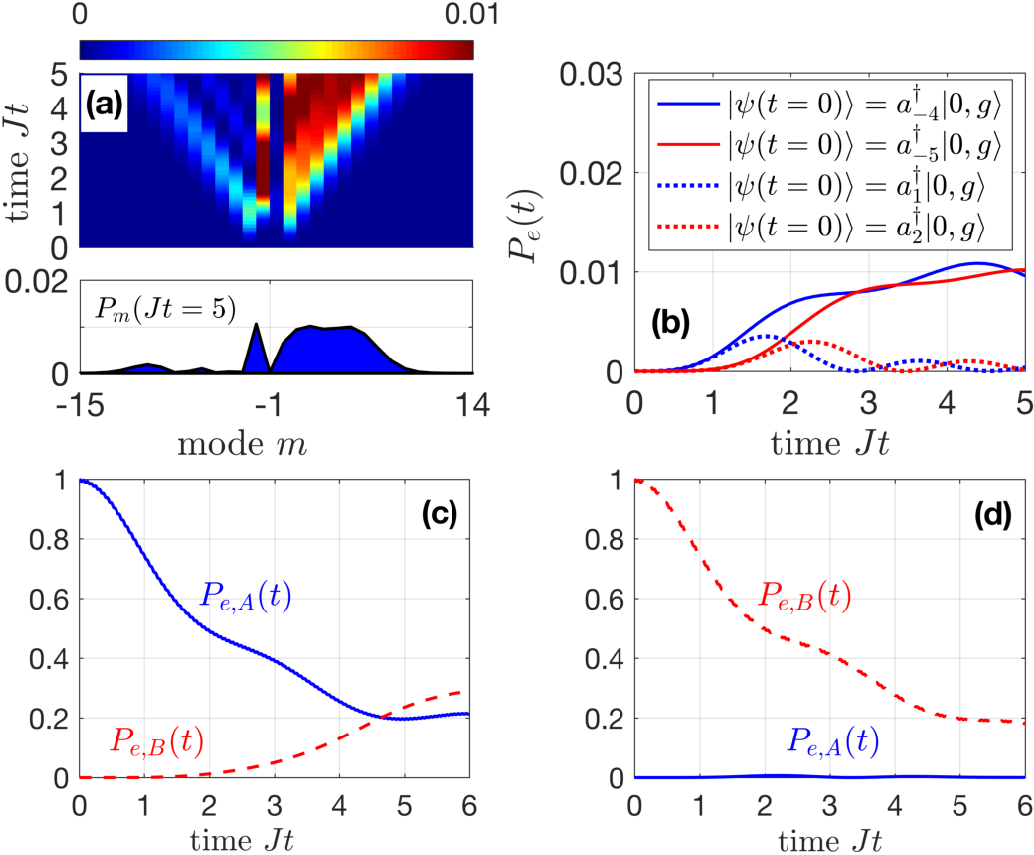}
\caption{(a) and (b) Dynamic evolutions of (a) $P_{m}(t)$ with initial state $|\psi(t=0)\rangle=|0,e\rangle$ and (b) $P_{e}$ with different initial states. The lower panel in (a) depicts the modal excitation profile at $Jt=5$. (c) and (d) Dynamic evolutions of atomic populations $P_{e,A}(t)$ and $P_{e,B}(t)$ of two separate effective giant atoms with initial states (c) $|\psi(t=0)\rangle=|0,e,g\rangle$ and (d) $|\psi(t=0)\rangle=|0,g,e\rangle$. We assume $g_{0}=0.1J$, $g_{N}=3J$, $\eta=2J$, and $\Delta=60J$ in (a) and (b) and assume $g_{0}=0.4J$, $g_{N}=10J$, $\eta=4J$, and $\Delta=100J$ in (c) and (d). Other parameters are $N=-3$, $\theta=\pi/2$, and $\gamma=0.01J$.}\label{fig3}
\end{figure}

\emph{Chiral quantum optics in the frequency dimension}.--Up to now, we have considered the case $\theta=0$, which simulates in the frequency dimension a conventional two-level giant atom. Now we turn to consider the case $\theta\neq0$, which can be readily achieved by tuning the phase of the coherent external field. In this case, the Hamiltonian (\ref{eq3}) becomes asymmetric in momentum space such that the spontaneous emission of the giant atom should be chiral~\cite{WXchiral}. In particular, the giant atom exhibits almost unidirectional emission when $\theta=\pi/2$ and $N=-3$, as shown by the modal excitation probability $P_{m}(t)=|a_{m}(t)|^{2}$ in Fig.~\figpanel{fig3}{a}. This phenomenon can be understood from the effective Hamiltonian of Eq.~(\ref{eq6}), which reads
\begin{equation}
\tilde{H}=H_{r}+\Big[|g\rangle\langle e|\Big(g_{0}a_{0}^{\dag}+g_{e,N}a_{N}^{\dag}e^{-i\theta}\Big)+\textrm{H.c.}\Big]
\label{eq8}
\end{equation}
if $\{\Delta_{e,e},\Delta_{e,N},\gamma\}\rightarrow0$. By performing the transformation $a_{k}=\sum_{m}a_{m}\textrm{exp}(-ikm)/\sqrt{2\pi}$, Eq.~(\ref{eq8}) becomes
\begin{equation}
\begin{split}
\tilde{H}'&=\int dk\omega_{k}a_{k}^{\dag}a_{k}+\frac{1}{\sqrt{2\pi}}\int dk\Big[\Big(g_{0}\\
&\quad\,+g_{e,N}e^{-i(kN+\theta)}\Big)|g\rangle\langle e|a_{k}^{\dag}+\textrm{H.c.}\Big]
\end{split}
\label{eq9}
\end{equation}
with $\omega_{k}=2J\cos{k}$ the dispersion relation of the lattice. Equation~(\ref{eq9}) shows a standard atom-field interaction but with an engineered momentum-dependent coupling. Clearly, the phase $\theta$ renders the coupling asymmetric in $k$ and thus mimics a synthetic gauge field threading the plaquette spanned by the atom-waveguide couplings. The gauge field allows the interaction to imprint a momentum kick on photons propagating toward left or right, similar to the Aharonov-Bohm (AB) effect. Once again, this result can be quantitatively understood from the model of giant-small atoms~\cite{SM}, where the excitation of the giant atom cannot be transferred to a small atom on its left side if $\theta=\pi/2$ and $N=-3$. We point out that the weak excitations of the left-side sites in Fig.~\figpanel{fig3}{a} arise from the finite-size effect of the lattice. As shown in Fig.~S2 of Ref.~\cite{SM}, the chiral profile tends to become more ideal with the increase of the lattice size. 

On the other hand, such a giant atom can hardly be excited by photons coming from the right side, as shown in Fig.~\figpanel{fig3}{b}. This thus provides the possibility of realizing cascaded interactions~\cite{cascade1,cascade2,cascade3} in the frequency dimension, with which the excitation of the present giant atom can be transferred to another such one but not vice versa. As an example, we assume that another giant atom (labeled as atom $B$) is coupled to the frequency lattice at sites $m=1$ and $m=4$, while the present one (labeled as atom $A$) is coupled to the lattice at sites $m=-3$ and $m=0$. Other parameters such as the effective coupling strengths are assumed to be identical for the two atoms. We plot in Figs.~\figpanel{fig3}{c} and \figpanel{fig3}{d} the evolutions of the atomic populations $P_{e,A}(t)$ of atom $A$ and $P_{e,B}(t)$ of atom $B$ when different atoms are initially excited. As expected, in the case of $\theta=\pi/2$, the excitation of atom $A$ can be partially transferred to atom $B$, whereas atom $A$ is always nearly in the ground state if atom $B$ is initially excited.  

\begin{figure}[ptb]
\centering
\includegraphics[width=8 cm]{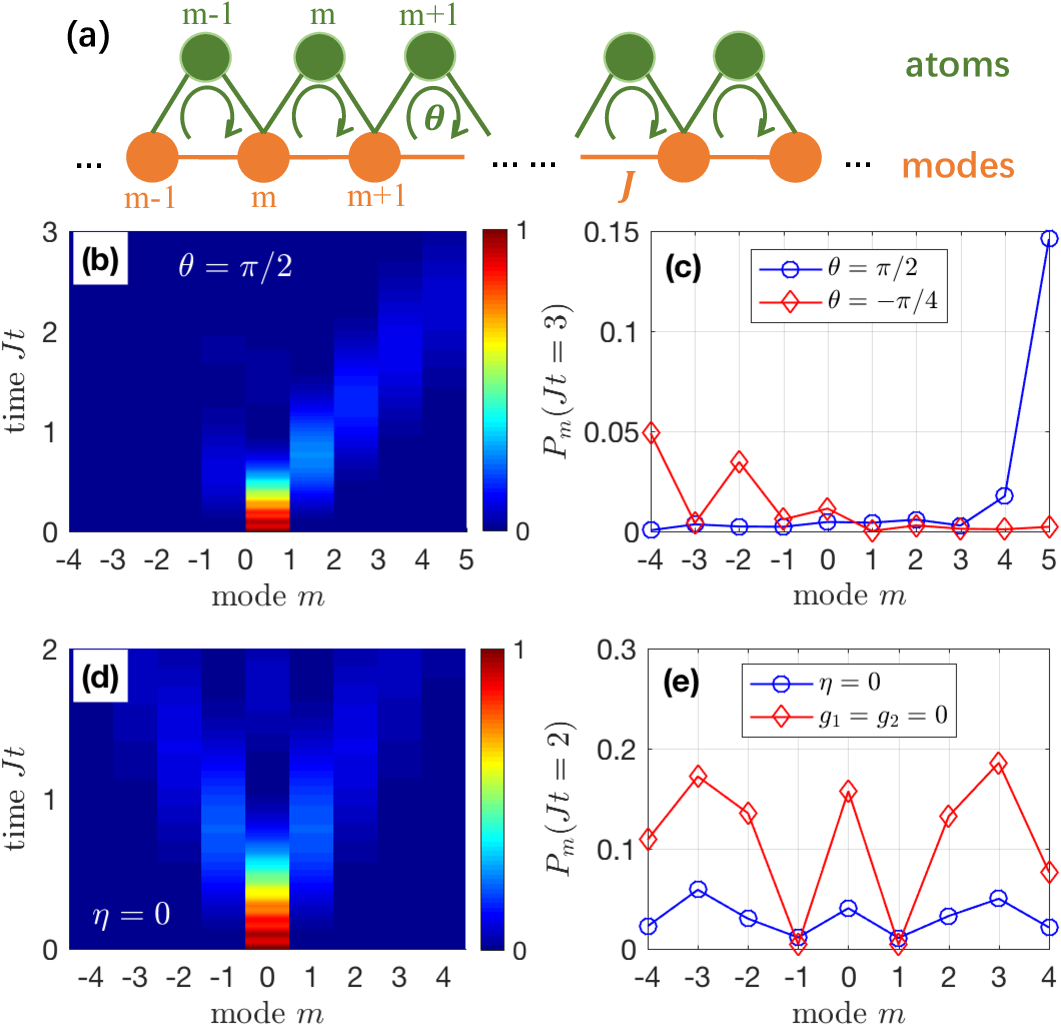}
\caption{(a) Schematic illustration of the effective sawtooth lattice. (b) and (d) Dynamic evolutions of $P_{m}(t)$ in the sawtooth lattice with (b) $\theta=\pi/2$ and (d) $\eta=0$. (c) and (e) Modal excitation profiles $P_{m}$ at specific moments with different parameters. Other parameters are $Q=4$, $g_{1}=J$, $g_{2}=10J$, $\eta=6J$, $\gamma=2J$, $\Delta=60J$, $\theta=\pi/2$, and $|\psi(t=0)\rangle=a_{0}^{\dag}|G\rangle$ unless indicated.}\label{fig4}
\end{figure}


Based on the chiral mechanism above, we now demonstrate that the effective giant atom can be used as a photonic AB cage~\cite{ABcage} that enables directional excitation transfer along the frequency dimension. As shown in Fig.~\figpanel{fig4}{a}, we consider an array of such $\Delta$-type atoms ($2Q+1$ atoms in total), each of which couples to two nearest-neighbor lattice sites with strengths $g_{1}$ and $g_{2}$, respectively, to form a ``sawtooth lattice'' in the frequency dimension. The effective sawtooth lattice is assumed to be homogenous, with the Hamiltonian $H_{s}=H_{r}+H_{a}+H_{\textrm{int}}$ given by
\begin{equation}
\begin{split}
H_{a}&=\sum_{m=-Q}^{Q}[-i\gamma|e\rangle_{m}\langle e|-(\Delta+i\gamma)|f\rangle_{m}\langle f|],\\
H_{\textrm{int}}&=\sum_{m=-Q}^{Q}[g_{1}|g\rangle_{m}\langle e|a_{m+1}^{\dag}+g_{2}|g\rangle_{m}\langle f|a_{m}^{\dag}\\
&\quad\,+\eta e^{i\theta}|e\rangle_{m}\langle f|+\textrm{H.c.}],
\end{split}
\label{eq10}
\end{equation}
where the subscript $m$ also labels the atom array. When $\theta=\pi/2$, as shown in Fig.~\figpanel{fig4}{b}, the initial excitation $|\psi(t=0)\rangle=a_{0}^{\dag}|G\rangle$ ($|G\rangle$ is the ground state of the whole system) exhibits a nearly directional transfer as expected. The transfer direction and velocity can be controlled by tuning the phase $\theta$ [see, e.g., the modal profiles in Fig.~\figpanel{fig4}{c}]. This is reminiscent of the real-space non-Hermitian sawtooth lattice~\cite{BiLonghi,DLSR}, where only the excitation components with specific wave vectors can propagate along the lattice without loss due to the combination of the AB effect and the non-Hermiticity. We stress that the intrinsic atomic dissipation plays a key role for observing the directional transfer (see Sec.~V of Ref.~\cite{SM} for more details). Moreover, as shown in Figs.~\figpanel{fig4}{d} and \figpanel{fig4}{e}, the excitation transfer becomes nearly symmetric if the external fields are turned off ($\eta=0$) or in the absence of the atoms ($g_{1}=g_{2}=0$). Owing to the atomic dissipation, however, the modal excitation probability in the former case is typically smaller than that of the latter one.


\emph{Conclusions}.--In summary, we propose in this Letter a feasible scheme for constructing giant atoms in a synthetic frequency dimension. Compared with real-space giant atoms, our scheme has several advantages: (i) While real-space giant atoms typically require large-scale waveguides or coupled-resonator arrays, our scheme provides a hardware-efficient platform for exploring and simulating giant-atom effects. (ii) Our scheme suggests the possibility of simulating higher-dimensional giant atoms~\cite{HDgiant} in systems with lower geometric dimensionality, which are easier to construct experimentally. (iii) The present scheme opens up new venues for manipulating the frequency of light and achieving chiral quantum optical effects in the frequency dimension. The scheme in this Letter can also be extended to involve other synthetic dimensions~\cite{OAM3}, with which one can manipulate more degrees of freedom of photons and phonons.   

\begin{acknowledgments}
We acknowledge helpful discussions with G.~C. La Rocca and Peng Zhang. Y.L. is supported by the National Natural Science Foundation of China (under Grants No. 12074030 and No. U1930402). A.F.K. acknowledges support from the Swedish Research Council (Grant No. 2019-03696), and from the Knut and Alice Wallenberg Foundation through the Wallenberg Centre for Quantum Technology (WACQT). J.H.W. acknowledges the Funding from Ministry of Science and Technology of China (No.~2021YFE0193500).
\end{acknowledgments}

\end{document}